\documentclass[aps,pra,twocolumn,superscriptaddress,showpacs]{revtex4}
\usepackage{amsmath,amssymb,latexsym}
\usepackage[english]{babel}
\usepackage{graphicx}

\def\sx{X}
\def\sy{Y}
\def\sz{Z}

\def\l{\mathbf{l}}

\def\i{{\bf i}}

\def\j{{\bf j}}

\def\k{{\bf k}}
\def\m{{\bf m}}
\def\0{{\bf 0}}

\newcommand{\bea}{\begin{eqnarray}}
\newcommand{\eea}{\end{eqnarray}}

\newcommand{\bal}{\begin{align}}
\newcommand{\eal}{\end{align}}
\def\bi{\begin{itemize}}
\def\ei{\end{itemize}}
\def\bc{\begin{center}}
\def\ec{\end{center}}
\def\E{{\cal E}}

\def\C{\hbox{$\mit I$\kern-.7em$\mit C$}}
\def\R{\hbox{$\mit I$\kern-.6em$\mit R$}}

\def\ket#1{|#1\rangle}
\newcommand{\one}{\mbox{$1 \hspace{-1.0mm}  {\bf l}$}}

\def\tr{\mathrm{tr}}
\def\ket#1{\left| #1\right>}
\def\bra#1{\left< #1\right|}

\newcommand{\proj}[1]{\ket{#1}\bra{#1}}

\begin{document}

\title{Purification to Locally Maximally Entangleable States}

\author{Tatjana Carle}
\affiliation{Institute for Theoretical Physics, University of
Innsbruck, Innsbruck, Austria}
\author{Barbara Kraus}
\affiliation{Institute for Theoretical Physics, University of
Innsbruck, Innsbruck, Austria}
\author{Wolfgang D\"ur}
\affiliation{Institute for Theoretical Physics, University of
Innsbruck, Innsbruck, Austria}
\author{Julio I. de Vicente}
\affiliation{Institute for Theoretical Physics, University of
Innsbruck, Innsbruck, Austria}

\begin{abstract}
Locally maximally entangleable states (LMESs) constitute a large set of multipartite states, containing for instance all stabilizer states. LMESs are uniquely characterized by $(2^{n}-1)$ phases, where $n$ denotes the number of qubits. We consider here those LMES whose phases are either $0$ or $\pi$ and present a multipartite entanglement purification protocol for arbitrary such states. In contrast to all previously known recurrence protocols this protocol uses a novel ingredient, which is required due to the quantum correlations contained in the various LMESs. We compare this scheme to previously known entanglement purification protocols and show that the direct purification performs better than previously known protocols.
\end{abstract}
\maketitle

\section{Introduction}

The focus on quantum entanglement has shifted from a puzzling phenomenon of fundamental interest to a valuable resource in the context of quantum information processing. Several applications of entanglement are known, ranging from quantum metrology over quantum communication and computation to (multiparty) security applications, see e.g. \cite{Me,GoTh,RaBr01,AmFa08}. In all these applications pure states, often distributed among several spatially separated parties, are required. Despite spectacular experimental progress, the creation and maintenance of multiparticle entangled states remains a difficult task. Noise and decoherence limit the local generation of states, while channel noise is a main obstacle when creating distributed entanglement.

Entanglement purification is known to offer a possible way to overcome these limitations \cite{BeBr}. Entanglement purification protocols were introduced for bipartite states \cite{BeBr}, where it was shown that despite significant noise, e.g. resulting from sending particles through noisy channels, several copies of noisy entangled states allow one to generate fewer copies with high fidelity. Later entanglement purification protocols were introduced and analyzed for certain multipartite entangled states, in particular graph states or stabilizer states \cite{GlVa,De96,DuAsBr,AsDuBr,KrDu,Go06, DuBr} and W-states \cite{MiBr}. The possibility to purify certain states with the help of bipartite or multipartite entanglement purification protocols ensures that these states can be generated or maintained with high fidelity, even in the presence of noise. This is important when using such states as resources for quantum information processing or other tasks. In addition, its also a question of fundamental interest for which types of quantum correlations a purification process is possible.

Moreover, studying certain classes of states and investigating their entanglement features and possible applications has proven a to be a successful approach towards the understanding of multipartite entanglement --a complex problem that is still far from being completely understood.
Graph states or stabilizer states can be viewed as such an example. They constitute a large class of multiparticle entangled states, including several interesting and highly entangled quantum states such as the Greenberger-Horne-Zeilinger state, codewords for error correcting codes or the 2D cluster state \cite{Br01}. The latter serves as a resource for measurement based quantum computation \cite{RaBr01,Br09}.

Here we introduce a novel entanglement purification protocol for an even larger class of multiparticle entangled states, the the so-called locally maximally entangleable states (LMESs) \cite{KrKr}, thereby paving a way for possible practical applications of such states in the presence of channel noise and decoherence. LMESs are defined as those states where all quantum information can be maximally washed out by coupling each particle to a local auxiliary qubit and are equivalent to those states that can be used for optimal local encoding of classical information. LMESs correspond (up to local basis change) to a coherent superposition of all basis states, where all coefficients have equal norm and therefore are characterized by a phase. This leads to a total of $(2^{n}-1)$ parameters describing such a $n$-qubit state. In this paper we concentrate on LMESs where all phases are either $0$ or $\pi$, i.e. all coefficients are either (+1) or (-1). In this sense LMESs can be viewed as a generalization of graph states \cite{He04,HeBr} or stabilizer states, where  however stabilizers are no longer simple tensor products of Pauli operators. For all such LMESs, we provide a so-called recurrence protocol that allows one to purify a mixed state with sufficiently high initial fidelity to any desired LMES. More precisely, we present protocols that operate on two identical copies and which are capable of increasing the fidelity, provided the initial fidelity is sufficiently high. The purification process is divided into several sub-protocols. Each sub-protocol provides purification only with respect to certain parties, and the different sub-protocols need to be combined to achieve an overall entanglement purification with respect to the desired state.

The approach is similar to the one presented for the purification of graph states \cite{DuAsBr,AsDuBr,KrDu,Go06}, and in fact the resulting purification maps with respect to diagonal coefficients in the LMES basis turn out to be equivalent. However, new ingredients not used so far in entanglement purification protocols are needed to achieve this goal. It turns out that controlled-not operations or local stabilizer measurements \cite{GlVa} are no longer sufficient, and new types of  parity measurements are required. In contrast to Bell state or graph states purification protocols, it is also not possible to bring initial states to some standard form via local depolarization, i.e. diagonal in a LMES basis (which we call LME diagonal) as we show in this article. Nevertheless, we demonstrate that entanglement purification is possible for a large class of noisy LMES.

This paper is organized as follows. In Sec.\ \ref{stalme} we introduce our notation and review graph states and LMESs. In Sec. \ \ref{SecPPLME} we introduce and analyze our entanglement purification protocol. We first focus on a specific class of LMESs, and then show how this approach can be used to purify arbitrary LMESs with phases $0$ and $\pi$. In Sec.\ \ref{secdep} we discuss depolarization to LME diagonal states. There we show the impossibility of local depolarization, and provide an example where depolarization may spoil purification. In Sec.\ \ref{examples} we analyze several examples and compare our direct multiparticle entanglement purification protocol with possible alternative approaches, where we show an advantage of our approach in certain parameter regimes. We summarize and conclude in Sec.\ \ref{seccon}.

\section{Stabilizer States, Locally Maximally Entangleable States}
\label{stalme}
In this section we introduce our notation and review some basics concerning stabilizer and locally maximally entangleable states (LMESs).

 \subsection{Notation}

 Throughout the manuscript we use the following notation. $X,Y,Z$ denote the Pauli operators and $\one$ denotes the unnormalized identity operator. The subscript of an operator (state) denotes the qubit the operator (state) is acting on (describing) resp., e.g. $X_1$ denotes the $X$ operator acting on qubit $1$ and $\rho_1$ denotes the reduced state of qubit $1$. Unless stated otherwise, a superscript of an operator denotes the exponent of the operator, e.g. $X^0=\one$, and $X^1=X$. Moreover, we will often us multi--indices, which we denote by bold letters, like ${\bf k}=(k_1,\ldots, k_n)$.

\subsection{Stabilizer states and Graph states}
A $n$--qubit stabilizer state, $\ket{\Psi}$, is uniquely defined via a set of $n$ commuting operators $\{S_1, ..., S_n\}$ in the Pauli group \cite{GoTh}.
The state $\ket{\Psi}$ is the unique eigenstate to eigenvalue $+1$ for all the operators $S_i$, i.e. $S_i\ket{\Phi}=\ket{\Phi}$ $\forall i$ iff $\ket{\Phi}=\ket{\Psi}$. The group generated by $\{S_1, ..., S_n\}$ is called stabilizer group of $\ket{\Psi}$. We will call the generators, $S_i$, the stabilizer of $\ket{\Psi}$. Since the eigenvalues of each of the Pauli operators are $\pm 1$ the eigenvalues of any stabilizer are also $\pm1$ and are both $2^{n-1}-$fold degenerate. Simple examples for stabilizer states are Bell states, e.g. the stabilizer of $\ket{\Phi ^{+}}=1/\sqrt{2}(\ket{00}+\ket{11})$ is generated by $\{S_1=\sx \otimes \sx, S_2=\sz \otimes \sz\}$.

Note that each stabilizer state is up to local Clifford operations equivalent to a graph state \cite{He04, HeBr}. Graph states can be associated to a mathematical graph $G=(V,E)$ consisting of a set of vertices $V=\{1, \ldots, n\}$ and edges, $E\subseteq \{(i,j), i,j\in V, i\neq j\}$. The graph state corresponding to a graph $(V,E)$ is $\ket{\Psi}=U_{ph} \ket{+}^{\otimes n}$ with $U_{ph}=\prod_{(i,j)\in E} U_{ij}$. Here, $U_{ij}=\proj{0}_i \otimes \one_j+ \proj{1}_i \otimes Z_{j}$ denotes the $2-$qubit $\pi-$phase gate. The neighborhood of qubit $k$, $N_{k}:=\{l\in V|(k,l)\in E\}$, is the set of all vertices which are connected by an edge to qubit $k$.
It is easy to see that the stabilizer of a $n-$qubit graph state is generated by $S_i= U_{ph} X_i U^{\dagger}_{ph}=X_i \bigotimes_{j\in N_i} Z_j$, for $i\in V$.
Note that the common eigenbasis of the stabilizer operators, $S_i$, is given by $\{\ket{\Psi_\k}=Z^{k_1}\otimes Z^{k_2}\otimes \ldots \otimes Z^{k_n}\ket{\Psi}\}_{k_i\in\{0,1\}}$, with $\ket{\Psi_\0}=\ket{\Psi}$. Obviously, all elements of this basis are stabilizer states and it is easy to verify that $S_i \ket{\Psi_\k}=(-1)^{k_i}\ket{\Psi_\k}$. Thus, the index $\k$ contains all the information about the eigenvalues of the stabilizers to which $\ket{\Psi_\k}$ belongs to.

A useful concept in the context of graph states is colorability. Qubits which are not connected among each other by a phase gate can be grouped into a set called color.
A graph state is said to be $k-$colorable if one can group the set of vertices into $k$ subsets (colors) such that there are no edges among qubits within one subset. For instance, the $3$--qubit GHZ state, $\ket{GHZ}=1/\sqrt{2}(\ket{000}+\ket{111})$ is up to local unitary operators equivalent to $U_{12}U_{13}\ket{+}^{\otimes 3}$ and is therefore $2$--colorable.
Note that local unitary equivalent graph states might have a different colorability and that the minimal number of required colors is in general not known.

As mentioned in the introduction multipartite entanglement purification protocols have been derived to purify to arbitrary graph states \cite{DuAsBr,AsDuBr,KrDu,DuBr}. Here, we briefly summarize this purification protocol. Let us consider as a target state a $2$--colorable $n$--qubit graph state, $\ket{\Psi}$. The aim is to transform $N$ copies of an input state $\rho$ into $\ket{\Psi}$ via LOCC. Since the set $\{\ket{\Psi_\k}\}_{\k\in \{0,1\}^{n}}$ forms an orthonormal basis, $\rho$ can be written as $\rho=\sum_{\k, \l} \lambda_{\k}^{\l} \ket{\Psi_{\k}}\bra{\Psi_{\l}}$. It has been shown that any input state can be transformed via LOCC into a state diagonal in a graph-state basis, $\rho_{out}=\sum_{\k} \lambda_{\k } \ket{\Psi_{\k}}\bra{\Psi_{\k}}$, without changing the fidelity $F=\bra{\Psi}\rho\ket{\Psi}$ \footnote{Note that in this depolarization process non of the diagonal entries (in the graph state basis) is changed.}. Since all states $\ket{\Psi_\k}$ are local unitary (LU) equivalent to each other, the aim of entanglement purification is to gain information about the index $\k$, i.e. the eigenvalues of the stabilizers. Similarly to the recurrence protocol for Bell state \cite{BeBr}, the main idea is to transfer some information about the eigenvalues of the stabilizers from the first copy of the input state, $\rho_1$ to the second copy, $\rho_2$. This is achieved via multilateral CNOT operations. A measurement on the second copy reveals this information, which allows to purify the remaining copy. Since the qubits belonging to the same color do not interact, the information about the eigenvalues of the corresponding stabilizer can be obtained simultaneously. Later on, it has also been shown how arbitrary $k$--colorable graph states can be purified by using $2$--colorable graph states as auxiliary states. The required $2$--colorable graph states can be obtained from the $k$--colorable graph states via LOCC. As we will see, the tools required for the purification of LMESs will be significantly different. In particular, we will derive an entanglement purification protocol where other operations than the CNOT gates are required to reveal the desired information.

\subsection{Locally Maximally Entangleable states}
\label{lmesec}

In this section we review some properties of the class of LME states for which we derive purification protocols \cite{KrKr}. In the following we will consider $n$--qubit states and denote by $V=\{1,\ldots,n\}$ the set of all qubits. As mentioned in the introduction, a state is called LMES if it is possible to attach (in a certain way) to each system qubit a local auxiliary qubit such that the whole information of the system is washed out. That is the reduced state of the $n$--system qubits is completely mixed after attaching the auxiliary qubits. It has been shown that a $n$--qubit state is LME iff it is local unitary $($LU$)$ equivalent to a state of the form \bea \ket{\Psi}=U_{1, \ldots, n} \prod U_{i_{k_1},\ldots, i_{k_l}} \ldots \prod U_i \ket{+}^{\otimes n},\label{lme}\eea
where $U_{i_{k_1},\ldots, i_{k_m}}=\one +(e^{i\phi}-1)\proj{1}^{\otimes m}$ denotes a phase gate which is acting on the $m$ qubits, $\{i_{k_1},\ldots, i_{k_m}\}$ respectively. We call such a phase gate a pure phase gate of order $m$.
Equation  (\ref{lme}) reveals a way to generate the state, namely by applying the appropriate phase gates to a product state $\ket{+}^{\otimes n}$.
Looking at this generation process one can see that graph states are a subclass of LME states.
They are exactly those states which arise if one uses only 2-qubit interactions with $\pi$ phases. Notice that LMESs can be used to generate an arbitrary state of an even larger class of so-called M--states, which are used in the context of classical simulation of quantum computation \cite{Ne}. It is straightforward to show that any $n$--qubit M--state can be obtained from a $n+1$--qubit LMES by performing a measurement on a single qubit.

Due to the definition of LMESs one can generate an orthonormal basis of the Hilbert space by applying independent local unitary operations to the state. In particular, for an LMES $\ket{\Psi}$ as given in Eq.(\ref{lme}) one can apply local $Z$ operations to generate an orthonormal basis $\{\ket{\Psi_\k}=Z^{k_1}\otimes Z^{k_2}\otimes \ldots \otimes Z^{k_n}\ket{\Psi}\}_{k_i\in0,1}$. Moreover, it is easy to verify that the group generated by the operators $S_i=U_{ph} X_i U_{ph}^\dagger$ stabilizes the LMES $\ket{\Psi}=U_{ph}\ket{+}^{\otimes n}$. Like in the case of stabilizer states the operators fulfill the conditions $S_i^2=\one, [S_i,S_j]=0$ and $ S_i=S_i^\dagger$, for any $i,j \in V$.
Note however, that in contrast to stabilizer states the stabilizer of a LMES are no longer necessarily elements of the Pauli group, but act non-locally on several qubits.
Analogously to the stabilizer states, it can be easily verified that $S_i\ket{\Psi_{\k}}=(-1)^{k_i}\ket{\Psi_{\k}}$. Hence, $\ket{\Psi_\k}$ is an eigenstate of $S_i$ to eigenvalue $(-1)^{k_i}$.

Similarly to graph states we introduce the notion of neighborhood and colorability for LMESs. A qubit $i$ is said to be in the neighborhood of qubit $j$ if $U_{ph}$ contains a pure phase gate acting on both of them. The neighborhood, i.e. the union of all neighbors, of qubit $j$ will be denoted by $N_{j}$. We group those qubits of a LMES which are non-interacting among each other in different sets $A=\{a_1,\ldots a_{I_A}\}$, $B=\{b_1,\ldots,b_{I_B}\}$, etc for some $I_A,I_B,\ldots \in \{1,\ldots, n\}$. That is, there is no interaction between any two qubits belonging to the same set, say  $A$. In order to ease the notation we denote by $A$, $B$, etc. both, the set of qubits belonging to that color as well as the color itself. A LMES is then called $k-$colorable if $k$ is the number of required colors.
Whenever a $k$--coloration of a LMES has been chosen it will be advantageous to divide the multi--index $\k$ into the $k$ multi--indices $\k_A, \k_B,\ldots$ corresponding to the different colors.
We will then for instance write $\ket{\Psi_{\k_A,\k_B,\ldots}}$ instead of $\ket{\Psi_\k}$ when referring to the state $\otimes_{i=1}^{I_A} Z_{a_i}^{k_{a_i}}\otimes_{i=1}^{I_B} Z_{b_i}^{k_{b_i}}\ldots \otimes_{i=1}^{I_L} Z_{l_i}^{k_{l_i}}  \ket{\Psi_{\textbf{0}}}$.

From now on we will restrict ourselves to $\pi$--phases, i.e. any pure phase gate is of the form $U=diag(1,1,\ldots,1,-1)$. Note that this choice is not only for the sake of mathematical simplicity but also because of physical reasons as this class seems to be the most natural generalization of graph states. We call a $k$--colorable LMES, $\ket{\Psi}=U_{ph}\ket{+}^{\otimes n}$, regular if the decomposition of $U_{ph}$ into pure phase gates contains only pure phase gates of order $k$. Note that this implies that all pure phase gates are acting on one qubit per color. An example of a regular $3$--colorable LMES is $U_{123}U_{234}U_{345} U_{456} \ket{+}^{\otimes 6}$ with $A=\{1,4\}$, $B=\{2,5\}$, and $C=\{3,6\}$. Note that for a regular LMES, the set of neighbors of qubits in some color, say $A$, $N_A= \bigcup_{a_i\in A} N_{a_i}$, contains all qubits which are not in $A$. That is $N_A= N \backslash A\equiv \bar{A}$.
Thus, for any regular LMES, $\ket{\Psi}=U_{ph}\ket{+}^{\otimes n}$, we have $U_{ph}=\prod_{a_i\in A} U_{a_i}^{(a_i,N_{a_i})}$, where $U_{a_i}^{(a_i,N_{a_i})}\equiv\ket{0}_{a_i}\bra{0}\otimes \one + \ket{1}_{a_i}\bra{1}\otimes U_{a_i}$ denotes the phase gate (not necessarily pure) which entangles qubit $a_i$ with all its neighbors \footnote{Note that we consider here a kind of minimal colorability, that is, if one qubit is for instance not entangled to the rest, then one would not color the qubit with a new color.}. For instance for the example above we have $U_2^{(2,\{1,3,4\})}=U_{123}U_{234}=\proj{0}_2\otimes \one+\proj{1}_2 \otimes U_{13}U_{34}$. We will also use that $U_{ph}=\sum_\k \ket{\k}_A \bra{\k} \otimes U_\k$, where $U_\k=U_{a_1}^{k_1} \cdots U_{a_{|A|}}^{k_{|A|}}$ is acting on the neighbors of $A$. Using that $U_{a_1}=U_{a_1}^\dagger$ it is easy to verify that the stabilizers $S_{a_i}$ are given by $S_{a_i}=U_{ph}X_{a_i}U_{ph}^\dagger= X_{a_i} \otimes U_{a_i}$. Moreover, since $U_{a_i}$ is not acting on any qubit in $A$ for any $i$, we have that $S_{a_1}^{j_1}\cdots S_{a_{|A|}}^{j_{|A|}}=(X_A)^{\j_A} \otimes U_\j$. In the example above we have for instance, $N_A=\{2,3,5,6\}$ and $U_{ph}=\sum_\k \ket{\k}_{\{1,4\}} \bra{\k} \otimes U_\k$, where $U_{00}=\one$, $U_{01}=U_{4}=U_{23}U_{35}U_{56}$, $U_{10}=U_1=U_{23}$ and $U_{11}=U_1 U_4= U_{35}U_{56}$ and $S_1=X_1\otimes U_{23}$.

\section{Purification Protocol}
\label{SecPPLME}

Using the same notation as before, our aim is to purify an arbitrary $n$--qubit state, $\rho$, to a regular $n$--qubit LMES, $\ket{\Psi}$. Without loss of generality we take $|\Psi\rangle=|\Psi_\textbf{0}\rangle$, which is considered to be the element of the LME basis with which $\rho$ has the largest overlap.
In the last subsection of this section we address the problem of purifying to arbitrary $\pi$--LMES.

As usual in recurrence protocols, one assumes an arbitrarily large number of noisy copies at one's disposal.
Since the parties are spatially separated from each other, they are restricted to LOCC operations.

Let $P_A$ denote the projector onto the subspace spanned by those basis states, for which $k_A=\0$, i.e. $P_A=\sum_{\k_{\bar{A}}} \proj{\Psi_{\k_A=\0,\k_{\bar{A}}}}$. First of all, we show how one can transform two copies of a state $\rho$ into one $n$--qubit state, $\tilde{\rho}$ with $\tr(P_A \tilde{\rho})\geq \tr(P_A \rho)$ in case $\tr(P_A \rho)$ was sufficiently large. That is we {\it purify color $A$}. Then we will combine the purification protocol of all colors in order to increase the fidelity, $F=\bra{\Psi}\rho\ket{\Psi}=\tr(P_A P_B \cdots P_N \rho )$.

We consider as our target state a $k$--colorable regular LMES, $\ket{\Psi}=U_{ph}\ket{+}^{\otimes n}$. Using that the states $\ket{\Psi_{\k}}=\sz^{k_1}\otimes \sz^{k_2}\otimes \ldots \otimes \sz^{k_n}\ket{\Psi}$ form an orthonormal basis, we express the input state in this basis,
$\rho=\sum_{\k, \l} \lambda_{\k}^{\l} \ket{\Psi_{\k}}\bra{\Psi_{\l}}$. We derive now an LOCC protocol which achieves the following mapping:
\bea \label{eq:Purif} \ket{\Psi_{\k_A,\k_{\bar{A}}}}\ket{\Psi_{\l_A,\l_{\bar{A}}}}\rightarrow \ket{\Psi_{\k_A,\k_{\bar{A}} \oplus \l_{\bar{A}}}} H^{\otimes |A|} \ket{\k_A\oplus \l_A},\eea

where the last $|A|$ qubits are auxiliary qubits, which will be measured in the $X$--basis. Only if all measurement outcomes are $+1$ the state will be kept. Otherwise the copy is discarded. Thus, this protocol enables one to measure the parity $\k_A\oplus \l_A$. Hence, in case the initial weight $\tr(P_A \rho)$ was large enough, the probability of being in the right subspace with respect to color $A$, i.e. $\k_A=\0$, after the first purification step, is increased.

Considering Eq.(\ref{eq:Purif}) it is easy to see that an arbitrary input state $\rho=\sum_{\k, \k^\prime} \lambda_{\k}^{\k^\prime} \ket{\Psi_{\k}}\bra{\Psi_{\k^\prime}}$ will be transformed to the unnormalized state $\sum \lambda_{\k_A,\k_{\bar{A}}}^{\k^\prime_A,\k^\prime_{\bar{A}}}\lambda_{\k_A,\l_{\bar{A}}}^{\k^\prime_A,\l^\prime_{\bar{A}}} \ket{\Psi_{\k_A,\k_{\bar{A}}\oplus\l_{\bar{A}}}}\bra{\Psi_{\k^\prime_A,\k^\prime_{\bar{A}}\oplus\l^\prime_{\bar{A}}}}$. In particular, for an LMES--diagonal state, $\rho=\sum_{\k} \lambda_{\k} \ket{\Psi_{\k}}\bra{\Psi_{\k}}$ we obtain $\sum \lambda_{\k_A,\k_{\bar{A}}}\lambda_{\k_A,\l_{\bar{A}}}\ket{\Psi_{\k_A,\k_{\bar{A}}\oplus\l_{\bar{A}}}}\bra{\Psi_{\k_A,\k_{\bar{A}}\oplus\l_{\bar{A}}}}$.
The whole purification protocol consists then of a sequence of purifications of all colors. Even though the purification of some color, $B$, introduces again some noise in color $A$, i.e. by purifying color $B$ one might decrease $\tr(P_A \rho)$, an overall increase of the fidelity, $F$ is achieved. In particular, the target state, $\ket{\Psi}$ is a fixed point of this protocol.

Let us now explain in detail how the purification protocol for one color, say color $A$, works.
\subsection{The purification protocol for one color for regular LMESs}
\label{1color}

Let us denote by $a_i^1, A^1$ ($a_i^2, A^2$) the qubits and the set of qubits belonging to color A of the first (second) copy resp. Then, the purification protocol for color $A$, denoted by ${\cal P}_A$, reads as follows.

\begin{figure}[h!]
\begin{center}
  \includegraphics[scale=0.42]{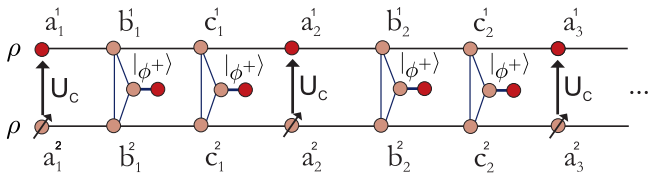}
  \caption{This graphic illustrates the protocol for a $3-$colorable LME state where color A is purified, see Sec.\ref{1color}.
  The target state is given by $\ket{\Psi_{\textbf{0}}}=\prod_{i=2}^{n-1} U_{i-1,i,i+1}\ket{+}^{\otimes n}$.
  The parties in color A entangle copy $1$ with copy $2$ by applying CNOT-gates.
  All other parties make a GHZ-measurement on one qubit of each copy and one qubit of a maximally entangled state, $\ket{\Phi^{+}}$.
  This measurement is depicted as a triangle.
Then a measurement in the $X-$basis is performed on the qubits of color A of the $2$nd copy. This measurement is displayed by an arrow pointing in the upper right direction.
The qubits of the remaining state are colored in a darker fashion (online: bright red).{\label{pr}}}.

\end{center}
\end{figure}

\bi \item[i)] All parties holding a qubit of color $A$ apply locally a CNOT--gate, $U_C=\proj{0}\otimes \one+ \proj{1}\otimes X$, between qubit $a_i^1$ and $a_i^2$. In total the unitary operation $\prod_{a_i\in A} U_C^{a_i^2,a_i^1}$ is applied.
\item[ii)] Each party holding a qubit in the neighborhood of $A$, $x\in N_A$, creates locally the maximally entangled two--qubit state, $\ket{\Phi^+}_{x^\prime,x^{\prime\prime}}$ and measures qubit $x^1,x^2$ and $x^\prime$ in the GHZ--basis, $\{X \otimes\one \otimes Z^{j}\ket{GHZ}, \one \otimes X \otimes Z^{j}\ket{GHZ}, \one \otimes \one \otimes X^{i} Z^{j}\ket{GHZ}\}_{i,j=0,1}$. Only those outcomes where one of the states $\one_{12} \otimes W_i\ket{GHZ}$ with $W_i\in \{X^j Z^k\}_{j,k=0,1}$ has been measured are kept and the operation $W_i^\ast$ is applied to qubit $x^{\prime \prime}$. Otherwise the copy is discarded.
\ei
As we will show below, after the first two steps the $2n$--qubit state $\ket{\Psi_{\k_A,\k_{\bar{A}}}}\ket{\Psi_{\l_A,\l_{\bar{A}}}}$ will be transformed into the $(n+|A|)$--qubit state $ \ket{\Psi_{\k_A,\k_{\bar{A}} \oplus \l_{\bar{A}}}} H^{\otimes |A|} \ket{\k_A\oplus \l_A}$ (we ignore here those qubits which have been measured in the GHZ--basis and therefore factorize). Note that to each party in color $A$ one of the last $|A|$ qubits belongs to (see Fig. \ref{pr} for an illustration). More precisely, qubit $a_i^2$ is after steps (i) and (ii) is in the state $H\ket{k_A^i\oplus l_A^i}$.
\bi
\item[iii)] All parties holding a qubit of color $A$ measure their qubits $a_i^2$ in the $X$--basis. They keep only those instances where the outcome $+1$ is obtained and discard the rest. Thus, only those copies, where $\k_A\oplus \l_A=\0$ are kept.
\ei

Before we show that this protocol achieves indeed the desired transformation, we want to stress that in contrast to existing recurrence protocols, we need here the projection onto the GHZ--state (step (ii)). That is, in contrast to the Graph state protocol, it is not enough to apply multilateral CNOT--gates. The reason for this is the different kind of quantum correlations contained in LMES. To make this statement more precise let us consider the simple example of the three--colorable regular LMES, $\ket{\Psi}=U_{123}\ket{+}^{\otimes 3}$. Considering the application of $U_C^{1,1^\prime}\ket{\Psi_\k}_{123}\ket{\Psi_\l}_{1^\prime 2^\prime 3^\prime}$ we obtain, $Z_1^{k_1l_1} U_{1^\prime 23}\ket{\Psi_\k}_{123}\ket{\Psi_\l}_{1^\prime 2^\prime 3^\prime}$. Thus, the CNOT gate entangles the two copies in such a way that no other local CNOT gate can disentangle them again keeping some information about the eigenvalues of $S_{a_i}$ in the second copy. Therefore, any subsequent measurement on the second copy, which is required to "project" the remaining copy onto the desired subspace, will destroy the entanglement contained in the first copy. Due to that, it is also not possible to consider a purification protocol, where more than just two copies are considered, and only CNOT--gates are required.

Let us now show that the protocol ${\cal P}_A$, i.e. the purification of color $A$, achieves the following mapping:
\bea \label{eq:Purif1} \ket{\Psi_{\k_A,\k_{\bar{A}}}}\ket{\Psi_{\l_A,\l_{\bar{A}}}}\rightarrow \delta_{\k_a,\l_A}\ket{\Psi_{\k_A,\k_{\bar{A}} \oplus \l_{\bar{A}}}}.\eea

Let us first investigate the projection onto the GHZ--states, $\one_{12}\otimes W_i \ket{GHZ}$, with $W_i\in \{X^j Z^k\}_{j,k=0,1}$. Denoting by $P^{b_i^1 b_i^2 b_i^3}_{GHZ}$ the projector onto the $GHZ$--state, we have $P^{b_i^1 b_i^2 b_i^3}_{GHZ} \ket{\Phi^+}_{b_i^3, b_i^{3^\prime}}=\frac{1}{2}\ket{GHZ}_{b_i^1 b_i^2 b_i^3}(\ket{0}_{b_i^{3^\prime}}\bra{00}_{b_i^1 b_i^2}+\ket{1}_{b_i^{3^\prime}}\bra{11}_{b_i^1 b_i^2})\equiv \frac{1}{2}\ket{GHZ}_{b_i^1 b_i^2 b_i^3} P^{b_i}$.
Note that $P^{b_i}\ket{\psi}_{b_i^1}\ket{\phi}_{b_i^2}=(\ket{\psi} \bigodot \ket{\phi})_{b_i^{3^\prime}}$, where $\ket{\psi} \bigodot \ket{\phi}$ denotes the Hadamard product between the states $\ket{\psi}$ and $\ket{\phi}$. That is $\ket{\psi} \bigodot \ket{\phi}=\sum_i \alpha_i \beta_i \ket{i}$ for $\ket{\psi}=\sum_i \alpha_i \ket{i}$ and $\ket{\phi}=\sum_i \beta_i \ket{i}$. As can be easily shown the projection onto the state $\one_{12}\otimes W_j \ket{GHZ}$ leads to $W_j^T P^{b_i}$. The local unitary operation is undone by applying $W_j^\ast$ (see step (ii)). Note that the probability of measuring one of the four states $\{\one_{12}\otimes W \ket{GHZ},$ with $W\in \{X^j Z^k\}_{j,k=0,1}\}$, i.e. the probability of successfully implementing the projector $P^{b_i}$ is larger or equal to $1/2$, as can be easily seen as follows. Let $\rho_i=\sum_{nm} r_{nm}\ket{n}\bra{m}$ denote the reduced state of particle $b_i$ of $\rho$. Then, the success probability, $p_{succ}=\tr(P^{b_i} \rho_i \otimes \rho_i P^{b_i})=\tr (\rho_i \bigodot \rho_i)=r_{00}^2+r_{11}^2$. Using that $\tr(\rho_i)=r_{00}+r_{11}=1$ we find $p_{fail}=1-p_{succ}=2r_{00}r_{11}$. Since $p_{succ}-p_{fail}=(r_{00}-r_{11})^2\geq 0$ we have $p_{succ}\geq 1/2$.

Let us in the following denote by $P^{N_{a_i}}$ the product of $P^{b_j}$ for all neighbors,$b_j$, of $a_i$. Note that $P^{N_{a_i}}\ket{\psi} U_{ph}\ket{+}^{\otimes |A|}=U_{ph}\ket{\psi}$, for arbitrary phase gates $U_{ph}$ and $|A|$--qubit states $\ket{\psi}$.

With all that it is now straightforward to prove Eq (\ref{eq:Purif1}). Combining steps (i) and (ii) we have:
\bea \ket{\Phi}=\bigotimes_{a_i\in A} P^{N_{a_i}}\bigotimes_{a_i\in A} U_C^{a_i^2, a_i^1} \ket{\Psi_{\k_A,\k_{\bar{A}}}}\ket{\Psi_{\l_A,\l_{\bar{A}}}}=\\ \nonumber
\bigotimes_{a_i\in A} P^{N_{a_i}}\sum_{\j_A}  X_{A_1}^{\j_A}\ket{\Psi_{\k_A,\k_{\bar{A}}}}_{A_1,\bar{A}_1} \proj{\j_A}_{A_2} Z_{A_2}^{\l_A} \\ \nonumber Z_{\bar{A}_2}^{\l_{\bar{A}}}\sum_{\m_A}\ket{\m_A}_{A_2}U_{\m_A}\ket{+}^{\otimes N-|A|}_{\bar{A}_2}.\eea Using now that both, $\bigotimes_{a_i\in A} P^{N_{a_i}}$ and $U_{\m_A}$ only act on the neighbors of $A_1$, $A_2$ resp. and that $ Z_{\bar{A}_2}^{\l_{\bar{A}}} \ket{\Psi_{\k_A,\k_{\bar{A}}}}=\ket{\Psi_{\k_A,\k_{\bar{A}}\oplus \l_{\bar{A}}}}$ we have
\bea \label{eq:Purif1a} \ket{\Phi}=
\sum_{\j_A}  (-1)^{\l_A\cdot \j_A} X_{A_1}^{\j_A}\otimes U_{\j_A} \ket{\Psi_{\k_A,\k_{\bar{A}}\oplus \l_{\bar{A}}}}\ket{\j_A}_{A_2}=\\ \nonumber
\sum_{\j_A} \ket{\Psi_{\k_A,\k_{\bar{A}}\oplus \l_{\bar{A}}}}\ket{\j_A}_{A_2} (-1)^{(\l_A\oplus \k_A)\cdot \j_A}=\\ \nonumber
\ket{\Psi_{\k_A,\k_{\bar{A}}\oplus \l_{\bar{A}}} }H^{\otimes |A|} \ket{\l_A\oplus \k_A}. \eea

Note that the first equality in Eq (\ref{eq:Purif1a}) is due to the fact that $X_{A_1}^{\j_A}\otimes U_{\j_A}=\prod_{i\in \{1,\ldots, |A|\}} S_{a_i}^{j_i}$, which implies that $X_{A_1}^{\j_A}\otimes U_{\j_A}\ket{\Psi_{\k_A,\k_{\bar{A}}\oplus \l_{\bar{A}}}}=(-1)^{\k_A\cdot \j_A}\ket{\Psi_{\k_A,\k_{\bar{A}}\oplus \l_{\bar{A}}}}$. Measuring now the qubits $a_i^2 \in A_2$ in the $X$--basis and keeping only those instances where all measurement outcomes are $+1$ leads to Eq (\ref{eq:Purif1}).

The intuition behind the subprotocols is the following. The projection on the $GHZ$--state applied to the neighbors of qubit $a_i$, combined with the application of the CNOT gate on the qubits $a_i^1, a_i^2$ for all $i$ maps the information about the parity $\k_A\oplus \l_A$ into the qubits in $A_2$, which are then measured. That is, some information about the first copy is mapped to the second, which is revealed by the measurement (see Eq (\ref{eq:Purif1})). This information gain leads to the purification of the state. As mentioned above the subprotocol ${\cal P}_A$ maps an arbitrary input state $\rho=\sum_{\k, \k^\prime} \lambda_{\k}^{\k^\prime} \ket{\Psi_{\k}}\bra{\Psi_{\k^\prime}}$ to the unnormalized state $\tilde{\rho}=\sum_{\k, \k^\prime} \tilde{\lambda}_{\k}^{\k^\prime} \ket{\Psi_{\k}}\bra{\Psi_{\k^\prime}}$, where
\begin{equation}\label{protocol}
\tilde{\lambda}_{\k_A,\k_{\bar{A}}}^{\k^\prime_A,\k^\prime_{\bar{A}}} = \sum_{\l_{\bar{A}},\l'_{\bar{A}}} \lambda_{\k_A,\k_{\bar{A}}}^{\k^\prime_A,\k^\prime_{\bar{A}}}\lambda_{\k_A,\l_{\bar{A}}\oplus\k_{\bar{A}}}^{\k^\prime_A,\l^\prime_{\bar{A}}\oplus\k'_{\bar{A}}} .
\end{equation}

In the particular case of a LMES--diagonal state, $\rho=\sum_\k \lambda_{\k}\proj{\Psi_\k}$, the output is the unnormalized LMES--diagonal state, $\tilde{\rho}=\sum_\k \tilde{\lambda}_{\k}\proj{\Psi_\k}$  with
\bea \label{Eq:lambda}\tilde{\lambda}_{\k_A,\k_{\bar{A}}}= \sum_{\l_{\bar{A}}} \lambda_{\k_A,\l_{\bar{A}}} \lambda_{\k_A,\l_{\bar{A}}\oplus\k_{\bar{A}}}.\eea
Note that this equation resembles the one obtain for the purification protocols for graph states even though the protocol itself is fundamental different. The reason for that is that in both cases the second copy of the state is used to obtain the information about the bit values $\k_A\oplus \l_A$ corresponding to parties in set A and in both cases only those instances where $\k_A\oplus \l_A=\0$ are kept.

In order to illustrate the subprotocol we consider the simple example of a $n=3m$--qubit $3-$colorable regular LMES $\ket{\Psi}=\prod_{i=2}^{n-1} U_{i-1,i,i+1}\ket{+}^{\otimes n}$, for $m$ being an arbitrary integer. In Figure \ref{pr} we depict the subprotocol ${\cal P}_A$ for this state. The stabilizers of color $A$ have the following form $\{S_{a_1}=X_{a_1}\otimes U_{b_1 c_1}, S_{a_2}=X_{a_2}\otimes U_{b_1 c_1} U_{c_1 b_2} U_{b_2 c_2}, S_{a_3}=X_{a_3}\otimes  U_{b_2 c_2} U_{c_2 b_3}U_{b_3 c_3},\ldots  S_{a_m}=X_{a_m}\otimes  U_{b_{m-1} c_{m-1}} U_{c_{m-1} b_m}U_{b_m c_m}\}$. As explained above, the information about $\k_A$ is transferred to the second copy by applying CNOT gates between the qubits in $A_1$ and $A_2$ and by performing the GHZ--measurements on all neighboring qubits, $b_i,c_i$, for $i=1,\ldots m$ (see Figure \ref{pr}). The information is then gained by measuring the qubits $a_i^2 \in A_2$.

\subsection{Total Purification Protocol for regular LMESs}

The total multipartite entanglement purification protocol is composed of the subprotocols, ${\cal P}_X$, for all color $X$. First ${\cal P}_A$ is applied to two copies of the initial state. Then, ${\cal P}_B$ is applied to two copies of the output state, etc. Note that the convergence of the protocol depends on the order of the applied subprotocols. For certain input states an adaptive scheme is more advantageous than an alternating application of the subprotocols, as we show in Sec.\ref{pc}.

Due to Eq.\ (\ref{protocol}) the new coefficients, $\tilde{\lambda}^{\k'}_{\k}$, obtained after applying one subprotocol, are non-linear functions of the coefficients $\lambda_{\k}^{\k'}$. Thus, it is in general hard to find conditions under which the purification protocol, which is composed of several subprotocols ${\cal P}_X$, converges to the desired state. Nevertheless, it is easy to see that a necessary condition for convergence is that the fidelity with the target state, $\lambda_{{\bf 0}}$, is sufficiently large. In Sec.\ref{examples} we determine the threshold for the fidelity for which the protocol still succeeds for several different states and noise models. However, one must keep in mind that, due to the complexity of multipartite entanglement, this single figure of merit cannot be the only relevant quantity (e.\ g.\ one can find pairs of states with the same fidelity with the target state but one can be purified while the other cannot).

\subsection{Purification Protocol for $\pi$--LMESs}

In this section we explain how arbitrary $\pi$--LMESs can be purified. Let us call a phase gate, $U_{ph}$, regular phase gate if the corresponding LMES, $U_{ph}\ket{+}^{\otimes n}$ is a regular LMES. First of all, we note that an arbitrary phase gate $U_{ph}$ can be decomposed into regular phase gates, where the number of qubits the regular phase gates are acting on can vary. For instance the gate $U_{ph}=U_{123}U_{23}$ is decomposed into the regular phase gates, $U_{123}$ and $U_{23}$. Second, note that two LMESs, $\ket{\Psi_1}=(U_{ph}^1)_{i_1,\ldots,i_k}\ket{+}^{\otimes k}$ and $\ket{\Psi_2}=(U_{ph}^2)_{j_1,\ldots,j_l}\ket{+}^{\otimes l}$, for $l,k\in V$ with $\{i_1,\ldots,i_k\} \subseteq \{j_1,\ldots,j_l\}$ can be transformed via LOCC into the LMES  $\ket{\Psi}=(U_{ph}^1)_{i_1,\ldots,i_l}(U_{ph}^2)_{j_1,\ldots,j_l}\ket{+}^{\otimes l}$. This can be easily achieved by applying the GHZ--measurement explained in step (ii) in Sec.\ref{1color} among all qubits in $\{i_1,\ldots,i_k\}$. Note that the condition $\{i_1,\ldots,i_k\} \subseteq \{j_1,\ldots,j_l\}$ is not a restriction since one can always add qubits to the set $ \{j_1,\ldots,j_l\}$ onto which $U_{ph}^2$ is acting trivially. Thus, in order to purify an arbitrary LMES, one would first decompose the phase gate into regular phase gates, purify each of the corresponding regular LMESs and then combine the regular LMESs to obtain the desired state. For certain states there are definitely more suitable protocols than the one outlined above. For instance, the state $\rho=x\proj{\Psi}+\frac{1-x}{8}\one$, with $\ket{\Psi}=U_{123}U_{23}\ket{+}^{\otimes 3}$ can be purified to $\ket{\Psi}$ as follows. First, measure the first qubit in the $Z$--basis. In case outcome $+1$ is obtained, i.e. the state of the first qubit is projected onto $\ket{0}$, the state is (up to normalization) $\rho=x\proj{\Psi_2}+\frac{1-x}{8}\one$, where $\ket{\Psi_2}=U_{23}\ket{+}^{\otimes 2}$. This state can then be purified to $\ket{\Psi_2}$, which can then be used to perform the GHZ--measurement between $\rho$ and $\proj{\Psi_2}$ to obtain $\rho^\prime=\rho\bigodot (\proj{\Psi_2})= x\proj{\Psi_1}+\frac{1-x}{8}\one$, with $\ket{\Psi_1}=U_{123}\ket{+}^{\otimes 3}$. Since $\ket{\Psi_1}$ is a regular LMES $\rho^\prime$ can be purified using the methods explained in the previous sections. In a last step the GHZ measurements on one copy of $\ket{\Psi_1}$ and one copy of $\ket{\Psi_2}$ are performed to obtain $\ket{\Psi}$.

\section{Depolarization}
\label{secdep}

In this section we study how one could bring an arbitrary state $\rho=\sum_{\k, \l} \lambda_{\k}^{\l} \ket{\Psi_{\k}}\bra{\Psi_{\l}}$ to its diagonal form in the LME basis $\rho_d=\sum_{\k} \lambda_{\k}\ket{\Psi_{\k}}\bra{\Psi_{\k}}$.
For LME states this cannot be achieved in general by a deterministic LOCC map as we will show in this section.
This is in contrast to the stabilizer states and graph states where the process can be implemented locally and where the depolarization is used as a first step in purification protocols since this simplifies the analysis. We will nevertheless show later how this can be done non-locally using LME states as a resource.

In order to depolarize an arbitrary state $\rho$ to an LMES-diagonal form, one can apply the map \bea \label{depmaptwirl} \E_{twirl}(\rho)=\E_n \circ \E_{n-1} \circ \ldots \E_1(\rho)\eea with $\E_i=\frac{1}{2} (\rho+S_i \rho S_i^{\dagger})$ and the $S_i$ are the stabilizers of the LME state $\ket{\Psi_{\textbf{0}}}$.
The map makes the off-diagonal terms vanish, since $S_i \ket{\Psi_{\k}}\bra{\Psi_{\l}} S_i^{\dagger}=(-1)^{k_i \oplus l_i} \ket{\Psi_{\k}}\bra{\Psi_{\l}}$ and thus, for $k_i \neq l_i$ the off-diagonal terms of $\rho$ and $S_i \rho S_i^{\dagger}$ cancel each other.
Notice that --unlike in the case of graph states-- the operations $S_i$ are non-local.

In the following we are going to show that the depolarizing map and physically reasonable variations of it cannot be implemented by deterministic LOCC already in the simplest case of the 3--qubit regular LME state of order 3, i.\ e.\ $|\Psi\rangle=U_{123}|+\rangle^{\otimes3}$. This raises a second question: depolarizing simplifies the analysis of the purification protocol but it is not necessary as the scheme can nevertheless be applied to any state. Therefore, is it always advisable to depolarize before applying the purification map (in the sense that this enlarges the set of states whose fidelity with the target state can be brought arbitrarily close to 1 by the protocol)? In the bipartite case it is known that depolarization can actually spoil purification, in the sense that distillable states are mapped to non-distillable states. Here we show that this is also the case for our multipartite protocol, where we provide
a state $\rho$ for which our protocol succeeds in purifying to the 3--qubit regular LME state of order 3 but fails for $\E_{twirl}(\rho)$.
Notice that, since the mathematical form of the purification map is the same, this conclusion could be extended as well to 2-colorable graph states.
In summary, the purification procedure to LME states must be applied to possibly undepolarized states since in this case they cannot in general be made diagonal in the corresponding LME basis by LOCC.
However, this needs not be regarded as a drawback as depolarizing does not necessarily contribute to the success of the purification process and there are states for which depolarization would actually spoil the protocol. We will conclude this section by discussing different ways to implement the depolarizing map (\ref{depmaptwirl}) at the expense of consuming entanglement.

\subsection{Are depolarizing maps in the LME basis locally implementable?}

In this subsection we consider the use of a depolarizing map $\E_{dep}$ to preprocess the noisy state before the application of the purification protocol. One possible way of depolarization using a probabilistic application of stabilizer operators was already outlined above, however due to the non-locality of the stabilizers the corresponding operations cannot be implemented by LOCC. This does, however, not exclude the possibility that other sequences of LOCC can achieve the desired goal.

The most reasonable assumption for $\E_{dep}$ is that for any input state the off-diagonal entries are suppressed and the diagonal entries remain equal.
Moreover, the map should be independent of the state to be depolarized as we are aiming to devise general purification procedures which assume no previous information on the form of the noisy states to be purified. In other words, for any $n$-qubit state $\rho$ it should hold
\begin{align}\label{depmap1}
\langle\Psi_{\k}|\E_{dep}(\rho)|\Psi_{\k}\rangle&=\langle\Psi_{\k}|\rho|\Psi_{\k}\rangle\;\forall\k,\nonumber\\
\langle\Psi_{\j}|\E_{dep}(\rho)|\Psi_{\k}\rangle&=0\quad \forall \j\neq\k.
\end{align}
Let
\begin{align}
A_{\k}&=|\Psi_{\k}\rangle\langle\Psi_{\k}|\nonumber\\
B_{\j\k}&=|\Psi_{\j}\rangle\langle\Psi_{\k}|+|\Psi_{\k}\rangle\langle\Psi_{\j}|\quad (\j<\k)\nonumber\\
C_{\j\k}&=i(|\Psi_{\j}\rangle\langle\Psi_{\k}|-|\Psi_{\k}\rangle\langle\Psi_{\j}|)\quad (\j<\k),
\end{align}
where the relation $\j<\k$ is to be understood to hold when $\j$ and $\k$ are mapped from binary to decimal basis (i.\ e.\ in lexicographical order). It is straightforward to check then that
\begin{equation}\label{depmap1basis}
\E_{dep}(A_{\k})=A_{\k},\;\E_{dep}(B_{\j\k})=0,\;\E_{dep}(C_{\j\k})=0\;\forall\j<\k.
\end{equation}
Since $\{A_{\k},B_{\j\k},C_{\j\k}\}$ is a basis of the real vector space of Hermitian matrices, the map $\E_{dep}$ fulfilling Eqs.\ (\ref{depmap1}) is then unique for density matrices and therefore $\E_{dep}=\E_{twirl}$. Now, if the corresponding Choi-Jamiolkowski (CJ) state associated to a map $\E$ is entangled, then the corresponding operation cannot be implemented by LOCC \cite{CiLe}. Therefore, our question can be decided by computing the CJ state associated to $\E_{dep}$ for different LME states following Eq.\ (\ref{depmaptwirl}) or Eqs. (\ref{depmap1basis}). In the simplest case of the 3--qubit regular LME state of order 3 (i.\ e.\ $S_1=X_{1}\otimes U_{23},S_2=X_{2} \otimes U_{13},S_3=X_{3}\otimes U_{12})$, one, more precisely, needs to compute the CJ state
\begin{equation}
E^{dep}_{1a1b2a2b3a3b}=\E_{dep}^a\otimes\one^b(|\phi^+\rangle_{1a1b}|\phi^+\rangle_{2a2b}|\phi^+\rangle_{3a3b})
\end{equation}
and check whether or not it is fully separable in the splitting $1|2|3$. After lengthy but straightforward calculation one finds that $||(E^{dep}_{123})^\Gamma||_{tr}=1.75$, where the superscript $\Gamma$ indicates partial transposition with respect to one of the subsystems 1, 2 or 3 (the state is invariant under exchange of these subsystems). Since $||E^\Gamma||_{tr}=1$ iff the state has positive partial transposition, the corresponding CJ state $E^{dep}_{123}$ is not fully separable and, therefore, $\E_{dep}$ cannot be implemented by deterministic LOCC in this case.

One may nevertheless wonder whether a wider class of maps $\E'_{dep}$ with less stringent conditions but which still output states diagonal in the LME basis is still amenable by deterministic LOCC. Since one of the most relevant quantities for the purification to be possible is the fidelity of the noisy state with the target state, one may relax conditions (\ref{depmap1}) to
\begin{align}\label{depmap2}
\langle\Psi_{\mathbf{0}}|\E'_{dep}(\rho)|\Psi_{\mathbf{0}}\rangle&=\langle\Psi_{\mathbf{0}}|\rho|\Psi_{\mathbf{0}}\rangle,\nonumber\\
\langle\Psi_{\j}|\E'_{dep}(\rho)|\Psi_{\k}\rangle&=0\quad \forall \j\neq\k,
\end{align}
for any $n$-qubit state $\rho$ \footnote{One could in principle allow instead for maps that do not decrease the fidelity with the target state, as it is not clear to us whether this fidelity (optimized over local unitaries) is an entanglement monotone.}. In addition to this, we further restrict $\E'_{dep}$ to leave invariant all states which are already diagonal in the LME basis.
We will show, however, that the only positive map (i.\ e.\ mapping positive semidefinite matrices to positive semidefinite matrices) fulfilling the above conditions is again $\E_{dep}=\E_{twirl}$ and, hence, nothing is gained with such a relaxation. To see this, notice that the aforementioned restrictions on the action of $\E'_{dep}$ boil down to
\begin{equation}
\E'_{dep}(A_{\k})=A_{\k}\;\forall\k
\end{equation}
and, together with the fact that $\E'_{dep}$ should be trace-preserving, to
\begin{equation}
\E'_{dep}(B_{\j\k})=\sum_{\mathbf{p}\neq\mathbf{0}}\alpha^{\j\k}_{\mathbf{p}}A_{\mathbf{p}}
\end{equation}
with $\sum_{\mathbf{p}}\alpha^{\j\k}_{\mathbf{p}}=0$ $\forall \j,\k$ and similarly for $C_{\j\k}$. Clearly then, for each $\j$ and $\k$ at least one of the real numbers $\alpha^{\j\k}_{\mathbf{p}}$ must be negative for some value $\mathbf{p}$ (if $\E'_{dep}(B_{\j\k})\neq0$). However, using linearity and for some fixed choice $\l$ we have that
\begin{equation}
\E'_{dep}(aA_{\mathbf{0}}+bA_{\l}+cB_{\mathbf{0}\l})=aA_{\mathbf{0}}+bA_{\l}+c\sum_{\mathbf{p}\neq\mathbf{0}}\alpha^{\mathbf{0}\l}_{\mathbf{p}}A_{\mathbf{p}}
\end{equation}
for any $a,b,c\in\mathbb{R}$ ($c>0$). Now, if $\alpha^{\mathbf{0}\l}_{\mathbf{p}}$ is negative for some $\mathbf{p}\neq\l$ the map is clearly not positive so $\alpha^{\mathbf{0}\l}_{\l}$ must be negative. However, for any such possible negative value one can choose $b$ and $c$ such that $b+\alpha^{\mathbf{0}\l}_{\l}c<0$ (i.\ e.\ the output of the map is not positive semidefinite) and still choose $a$ such that $ab-c^2>0$ (i.\ e.\ the input to the map is positive semidefinite). Thus, one sees that in this case the map is also not positive. Therefore, the only possibility is $\E'_{dep}(B_{\j\k})=0 \;\forall \j\neq\k$ and obviously the same argument leads to $\E'_{dep}(C_{\j\k})=0 \;\forall \j\neq\k$. Thus, this leads to $\E_{dep}$, which cannot be implemented by LOCC in general for LME states.

\begin{figure}[h!]
\begin{center}
  \includegraphics[scale=0.46]{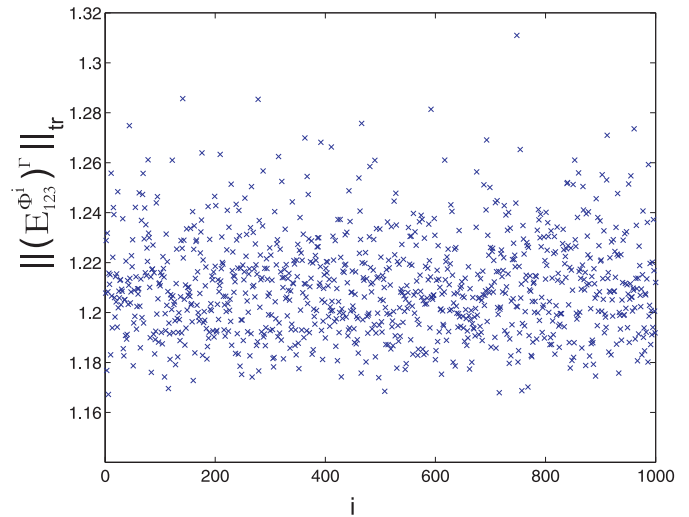}
  \caption{$||(E^{\Phi^{i}}_{123})^\Gamma||_{tr}$ for randomly generated maps $\Phi^{i}_{dep}$ $(1 \leq i \leq 1000)$. In all cases it holds that $||(E^{\Phi^{i}}_{123})^\Gamma||_{tr}>1$, indicating that the CJ state is entangled and, therefore, that the corresponding map cannot be implemented by LOCC.}\label{randommap}
\end{center}
\end{figure}

For the sake of completeness, let us finally consider the most general class of maps that output states diagonal in the LME basis preserving the fidelity with the target state. That is, those maps $\Phi$ satisfying just Eqs.\ (\ref{depmap2}) (dropping the condition that they leave invariant states which are already LMES--diagonal). We now have that
\begin{align}
\Phi_{dep}(A_{\mathbf{0}})&=A_{\mathbf{0}}\nonumber\\\label{depmap3}
\Phi_{dep}(A_{\k})&=\sum_{\mathbf{m}\neq\mathbf{0}}P_{\k\mathbf{m}}A_{\mathbf{m}}\;\forall\k\neq0
\end{align}
with P a (right) stochastic matrix (i.\ e.\ with all entries non-negative and $\sum_{\mathbf{m}}P_{\k\mathbf{m}}=1$ $\forall\k$) due to trace preservation. For the same reason as above, one can see that $\Phi_{dep}(B_{\j\k})=\Phi_{dep}(C_{\j\k})=0 \;\forall \j\neq\k$ must hold (since for any $0\leq P_{\l\mathbf{m}}\leq1$ and any $\alpha^{\mathbf{0}\l}_{\mathbf{m}}<0$ one can choose $b$ and $c$ such that $P_{\l\mathbf{m}}b+\alpha^{\mathbf{0}\l}_{\mathbf{m}}c<0$ and still choose $a$ such that $ab-c^2>0$). Hence, the support of $\Phi_{dep}$ is in the subspace spanned by the $\{A_{\k}\}$ and without loss of generality it can be considered as a composition $\Phi_{dep}=\Phi_P\circ\E_{dep}$ where $\Phi_P$ acts as given by Eqs.\ (\ref{depmap3}) for some choice of stochastic matrix $P$. This means that the CJ state for any $\Phi_{dep}$, $E^\Phi$, is given by
\begin{equation}
E^\Phi_{1a1b\cdots nanb}=\Phi_P^a\otimes\one^b(E^{dep}_{1a1b\cdots nanb}),
\end{equation}
but this does not necessarily imply that all $\Phi_{dep}$ cannot be implemented by LOCC since $\Phi_P$ for some choice of P might break the entanglement of the CJ state of $\E_{dep}$. However, for the 3--qubit regular LME state of order 3 the $P$ we have found decreasing the entanglement of $E^{dep}_{123}$ the most is still such that $||(E^{\Phi}_{123})^\Gamma||_{tr}=1.165$, which implies that $E^{\Phi}_{123}$ is nevertheless entangled. Moreover, we have generated 1000 maps $\Phi_{dep}$ with random choices for $P$ and found that they could not be implemented by LOCC  as the corresponding CJ state remains entangled (see Figure \ref{randommap}). Therefore, even in the most general case of depolarizing maps that just preserve the fidelity with the target state, there is strong numerical evidence indicating that no such map can be implemented by LOCC for the simplest case of the 3--qubit regular LME state of order 3.

\subsection{Depolarization may spoil purification}

As stated above, in many purification protocols a depolarization to the basis corresponding to the target state is performed before the actual protocol is applied.
This simplifies the mathematical analysis and one might think that this is favorable for the convergence of the purification process.
By convergence we mean that the target state can be reached up to any given accuracy by applying the protocol sufficiently often.
However, in this subsection we construct a counterexample which shows that off-diagonal elements can improve the threshold of convergence. Similar results are known in the bipartite case.
Consider, for instance, the case of purifying noisy states $\rho=\sum\lambda_{ijk}^{lmn}|\Psi_{ijk}\rangle\langle\Psi_{lmn}|$ to the $3-$qubit $3-$colorable state $|\Psi_{000}\rangle=U_{123}|+\rangle^{\otimes3}$. Using Eq.\ (\ref{protocol}), one can identify conditions on the off-diagonal entries $\{\lambda_{ijk}^{lmn}\}$ such that they add to $\tilde{\lambda}_{\textbf{0,0}}$ but do not add to the value of the other diagonal elements no matter which color is being purified, thus obtaining a larger fidelity with the target state after one iteration of the protocol than if the state was depolarized. This happens to be the case if it holds that $\lambda_{000}^{011}=-\lambda_{001}^{010}\in\mathbb{R}$, $\lambda_{000}^{101}=-\lambda_{001}^{100}\in\mathbb{R}$ and/or $\lambda_{000}^{110}=-\lambda_{010}^{100}\in\mathbb{R}$ and the rest of the off-diagonal entries are set to zero (up to the hermiticity condition $\lambda_{ijk}^{lmn}=(\lambda_{lmn}^{ijk})^\ast$). Therefore, one can find the desired counterexample by taking a diagonal state for which our purification protocol does not converge and populating strongly the off-diagonal elements as given above (as much as positive semidefiniteness allows). For instance, consider the state $\rho(f)$ with $\lambda_{\textbf{0,0}}=f$, $\lambda_{\textbf{i,i}}=(1-f)/7$ for $\i\neq\textbf{0}$ and
$\lambda_{000}^{011}=-\lambda_{001}^{010}=\lambda_{000}^{101}=-\lambda_{001}^{100}=\lambda_{000}^{110}=-\lambda_{010}^{100}=0.02$ (with the other off-diagonal entries up to hermiticity being zero). Notice that $\rho(f)$ is positive semidefinite when $0.01\lesssim f\lesssim0.72$. Moreover, if $f\gtrsim0.6503$, the purification protocol converges. However, the twirled state $\E_{twirl}(\rho(f))$ can be purified only if $f\gtrsim0.6507$.
Although the difference is small, this is enough to prove that it is not always more advantageous to depolarize the state before the purification protocol.

\subsection{Depolarization with nonlocal operations}

\begin{figure}[h!]
\begin{center}
  \includegraphics[scale=0.33]{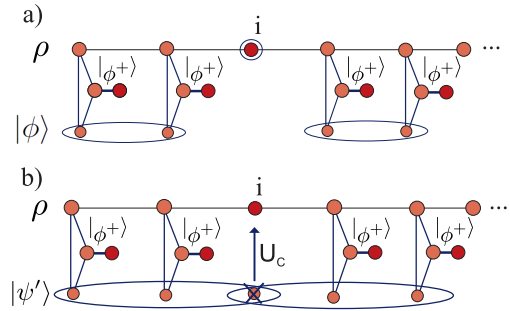}
  \caption{
  In this graphic we illustrate two methods to realize the CPM $\E_{i}=\frac{1}{2}(\rho+S_i \rho S^{\dagger}_i)$ of the depolarization process given by Eq.(\ref{depmaptwirl}) where $\ket{\Psi}=\prod_{i=2,4,6...} U_{i-1,i,i+1}\ket{+}^{\otimes n}$.
  Part a) depicts how $S_i$ is implemented using the auxillary state $\ket{\phi}$ (see main text).
   The measurement in the GHZ basis is pictured as the triangle which acts on $\rho$, the auxillary state, and one qubit of a maximally entangled state.
  The circle marks that a $X$ operator is applied. In order to realize the CPM $\E_i$ the parties choose randomly to apply this transformation or not.
  Part b) shows how $\E_i$ is implemented by employing the LME state $\ket{\psi'}$ and tracing out the qubit which is marked with a cross.}.{\label{dep_gr}}
\end{center}
\end{figure}

We have shown above that the map $\E_{twirl}$ given by Eq.\ (\ref{depmaptwirl}) that depolarizes to an LMES-diagonal state does not belong in general to the class of LOCC operations. However, this map can be implemented with certain probability of success at the expense of consuming some entanglement the parties might have initially at their disposal, or have generated from several noisy copies of their state with the help of entanglement purification protocols for different target states. One way to achieve this is the following. For any map $\E_i$ the parties decide randomly to apply $S_i$ or not.
This leads to $\E_i=\frac{1}{2} (\rho+S_i \rho S_i^{\dagger})$.
 The implementation of $S_i=\sx^{i}\otimes U_{i}$ can be easily done by applying the GHZ-measurement that we have considered in our purification protocol (cf.\ Sec.\ref{1color}), which succeeds with non-zero probability.
In particular, the $\sx$ of any $S_i$  can be applied locally by the $ith$ party while the phase gate is implemented by using an entangled state $\ket{\phi}=U_{i}\ket{+}^{\otimes m}$ shared among the $m$ parties holding the qubits in $N_i$ (to ease the notation we we do not explicitly write the dependence of $|\phi\rangle$ on the LME basis). This is because $S_i\ket{\Psi_{\k}}=\sx_{i}\otimes P^{N_i}\ket{\phi}\ket{\Psi_{\k}}$, where, as before, $P^{N_i}$ denotes the product of GHZ projections on all neighbors of $i$. Graphically this is shown in Figure \ref{dep_gr} $a)$. Notice that $U_{i}$ does not need to be pure and, hence, the $m$ parties holding $|\phi\rangle$ do not need to be fully entangled.   
Moreover, the phase gates which generate the auxiliary state $\ket{\phi}$ are of one order less than phase gates generating the LME basis in which the state is going to be diagonalized.
For regular LME states generated by phase gates of order $3$, $U_{i}$ consists therefore of two qubit phase gates only and the state $\ket{\phi}$ is then a (not necessarily connected) graph state for which purification protocols already exist.

The map $\E_{twirl}$ can also be implemented by consuming pure LME states $|\psi'\rangle$ of the same order as the one that corresponds to the basis in which the state is going to be depolarized. In order to implement the map $\E_i$ one has to use an auxiliary state $\ket{\psi'}=(\proj{0}_i\otimes \one+\proj{1}_i \otimes U_{i})\ket{+}^{\otimes m+1}$ and apply the CNOT between the qubit $i,i'$  and $P^{N_i}$ on the neighbors to obtain $\E_i(\rho)=\tr_{i'}(U_C^{i i'} P^{N_i}(\rho\otimes\proj{\psi'})P^{N_i} U_C^{i i'}).$
Here the primed index $i'$ is referring to the auxiliary state $\ket{\psi'}$. Figure \ref{dep_gr} $b)$ displays the procedure for one stabilizer $S_i$.

\section{Examples and Comparison}
\label{examples}

In this section we demonstrate the performance of our protocol by considering states, which are subjected to local as well as global noise. Then, we compare the direct purification protocol to indirect purification protocols, which, as we will show, can also be employed to purify to certain LMESs.

\subsection{Pauli channels}
\label{pc}

\begin{figure}[h!]
\begin{center}
  \includegraphics[scale=0.47]{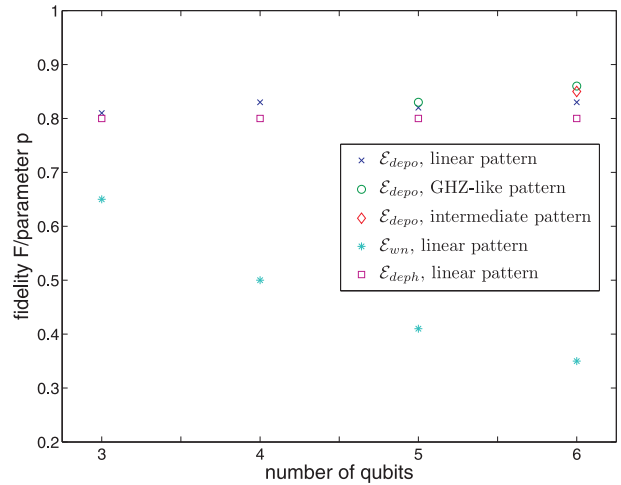}
  \caption{ In this figure we show the noise threshold $p$ of the depolarizing and dephasing channel given in Eq.(\ref{pd}) and Eq.(\ref{pp}) and the minimal required fidelity for the purification process for a state subject to global white noise of Eq.(\ref{wn}).
  We consider $3-$colorable LME states of $3,4,5$ and $6$ qubits which are generated by $3-$qubit phase gates.
            The interaction pattern were given by $U_{123}\ket{+}^{\otimes 3}$, $U_{123}U_{234}\ket{+}^{\otimes 4}$ for $3$ and $4$ qubits respectively $(\times)$.
            For the $5-$qubit state we have two possible patterns, a linear one $U_{123}U_{234}U_{345}\ket{+}^{\otimes 5}$ $(\times)$ and a GHZ-like one $U_{123}U_{124}U_{125}\ket{+}^{\otimes 5}$ $(\circ)$.
            In case of the $6-$qubit state one has three different patterns, a linear one $U_{123}U_{234}U_{345}U_{456}\ket{+}^{\otimes 6}$ $(\times)$, an intermediate one $U_{134}U_{235}U_{234}U_{346}\ket{+}^{\otimes 6}$ $(\diamond)$ and a GHZ-like one
            $U_{123}U_{124}U_{125}U_{126}\ket{+}^{\otimes 6}$ $(\circ)$.
            For the dephasing channel $(\square)$ the pattern does not strongly influence the threshold.
            For the global white noise channel $(*)$ we show the fidelity for a target LME state with a linear pattern.}.{\label{dep}}
\end{center}
\end{figure}

As examples of the noise model we consider first the local depolarizing as well as the local dephasing channel. The CPMs describing the local channels are given by $\E(\rho)=\bigotimes_i \E_i(\rho)$ where $\E_i$ is either
\bea \E_i(\rho)=p \rho+ \frac{(1-p)}{4}(\rho+\sx_{i} \rho \sx_{i}+ \sy_{i} \rho \sy_{i}+\sz_{i} \rho \sz_{i})\label{pd}\eea for the depolarizing channel or
\bea \E_i(\rho)=p \rho+ \frac{(1-p)}{2}(\rho+\sz_{i} \rho \sz_{i})\label{pp}\eea for the dephasing channel.
In the following we will use $\E_{depo}(\rho)(\E_{deph}(\rho))$ whenever we refer to the depolarizing $($dephasing$)$ channel.
We determined numerically the smallest value of $p$ for which the state still converges to the target state.
Note that the convergence depends on the sequence of subprotocols we choose for the purification. Moreover, the most advantageous sequence depends on the input state.
However, since we consider the case where one does not have any information about the noisy input state we investigate the examples for a fixed sequence.
We consider up to $6-$qubit $3-$colorable LME states with colors $A,B,C$. We iteratively apply the order $ABC-CAB-BCA$. This sequence seems to perform better than just repeating the sequence $ABC$ for a generic input state as the qubits in $C$ get more noisy due to the backaction from the subprotocols purifying $A$ and $B$. The states we investigate are given by $U_{123}\ket{+}^{\otimes 3}$ and $U_{123}U_{234}\ket{+}^{\otimes 4}$ for the $3-$ and $4-$qubit examples. For the $5$ and $6$ qubits we choose $U_{123}U_{234}U_{345}\ket{+}^{\otimes 5}$, $U_{123}U_{124}U_{125}\ket{+}^{\otimes 5}$ and $U_{123}U_{234}U_{345}U_{456}\ket{+}^{\otimes 6}$, $U_{134}U_{235}U_{234}U_{346}\ket{+}^{\otimes 6}$, $U_{123}U_{124}U_{125}U_{126}\ket{+}^{\otimes 6}$. The results we obtain for the depolarizing and dephasing channel (see Figure \ref{dep}) suggest that there are three crucial properties which influence the tolerated noise level, namely the interaction pattern of the target state,
the colorability of the target state and the number of qubits of the state.
The numerical results for the case of $3,4,5$, and $6$ qubits for different states are shown in Figure \ref{dep}.
For the depolarizing channel states with a {\it linear pattern}, $\prod_{i=2}^{n-1} U_{i-1,i,i+1}\ket{+}^{\otimes n}$, seem to be more robust than the ones with $GHZ-$like interactions, e.g. $\prod_{k=3}^{n}U_{12k}\ket{+}^{\otimes n}$. The noise level does not strongly depend on the number of qubits.
Note that similar results have been obtained for graph states \cite{AsDuBr}.
Note further, that the threshold for $p$ considering the depolarizing channel is close to the threshold of the dephasing channel even though the depolarizing channel outputs states which are non-diagonal and the dephasing channel outputs diagonal ones.
This supports our previous observation, that it is not drastically advantageous for the purification process to have LMES--diagonal states.

Let us now discuss how the colorability influences the noise thresholds. To this end we compare the tolerated noise level for the depolarizing channel acting on the considered $3-$colorable LMESs with the one obtained for the $2$--colorable states, i.e. linear cluster states and GHZ states in \cite{AsDuBr}. We find that the $3-$colorable states are more sensitive to noise.
Note that this result is not very surprising, since, as mentioned above, the purification of one color adds additional noise to the other colors. Thus, if there are many colors in the state we need more rounds until one color is purified for the second time. Hence, it is more often subjected to noise before it is purified again.
From a physical point of view one might argue that a state with higher colorability is more connected and therefore the errors introduced by the channels spread
more easily.

As a last example we also consider the influence of global noise. More precisely, we consider global white noise described by the CPM, ${\cal E}_{wn}(\rho)=x \rho+ \frac{1-x}{2^n}\one$, with $n$ denoting the number of qubits (see also Eq. (\ref{wn})) acting on $3$--colorable LMESs, $\ket{\Psi}=\prod_{i=2}^{n-1} U_{i-1,i,i+1}\ket{+}^{\otimes n}$. We calculate numerically the threshold for the fidelity, $f=\bra{\Psi}{\cal E}(\proj{\Psi})\ket{\Psi}$ such that the protocol still converges to the target state $\ket{\Psi}$. The numerical result for $3,4,5$ and $6$ qubit states is shown in Figure \ref{dep}, from which we can see how the threshold fidelity scales with the number of qubits in the state.

\subsection{Comparison to other purification protocols}

One way to generate a desired multipartite state is to generate maximally entangled bi--partite states shared between one party and all the others and to use teleportation to distribute the multipartite state. Another approach would be to purify states, which can be used to generate (via LOCC) the desired LMES.
For instance, we will show that it is possible to purify to certain LMESs by first purifying certain graph states and then recombining them in order to obtain the desired LMES. If the performance of such a scheme would be better than the multipartite entanglement purification scheme, the later would loose its practical relevance. The aim of this section is therefore to compare the LME-protocol with previously known purification protocols.  We will consider some example to show that the LME--purification protocol outperforms the graph state purification protocol as well as the bipartite purification protocol.

Let us consider the $6-$qubits linear LMES $\ket{\Psi}_6=U_{123}U_{234}U_{345}U_{456}\ket{+}^{\otimes 6}$. This state has the property that by measuring some of the qubits (of several copies), one can obtain certain graph (bi--partite) states, such that those states can be recombined via LOCC to again obtain the LMES. Thus, considering certain noise, one possibility to purify to the LMES would be to first apply the measurements to obtain the graph (bi--partite) states; purify them via the previously known entanglement purification protocols and then recombine them to the LMES. In order to show how these schemes compare to the direct LME--purification protocol we consider states which correspond to the output of a global white noise channel,
\bea \label{eq_whitenoise}\E_{wn}(\proj{\Psi})=x\proj{\Psi}+\frac{1-x}{2^{n}} \one \label{wn}\eea
where $\ket{\Psi}=\ket{\Psi}_6$ and $x=f-\frac{1-f}{2^{n}-1}$ with $f=\bra{\Psi} \E(\proj{\Psi}) \ket{\Psi}$ denoting the fidelity of the output state and $n=6$ is the number of qubits.

Before we explain several different schemes to purify to $\ket{\Psi}$ indirectly, we state the main result of this section. We consider the situation where an arbitrary number of copies of the state given in Eq. (\ref{eq_whitenoise}) is available and determine numerically the fidelity for which the various purification protocols succeed. Applying the LME--purification protocol presented here, we find that the protocol converges to $\ket{\Psi}$ as long as the initial fidelity, $f$, fulfills $f\geq0.345$.
However, purifying the graph states which are needed to generate the LME state would require a fidelity of $f\geq 0.349$. Using a bipartite strategy one requires a fidelity of $f>0.5$ \cite{BeBr}. Therefore, using the protocol we introduced here, one can purify states which are too noisy to be purified with any other strategy.

\begin{figure}[h!]
\begin{center}
  \includegraphics[scale=0.41]{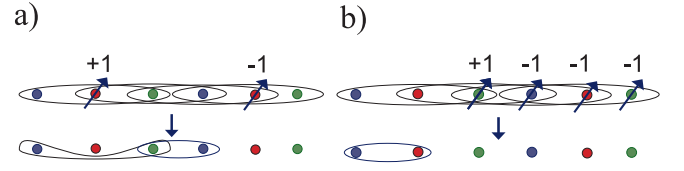}
  \caption{This figure illustrates how one can obtain graph (bi-partite) states from LME states by local $Z-$measurements.
  The input state is given by $\E(\proj{\Psi})$ from Eq.(\ref{wn}) with $\ket{\Psi}_6=U_{123}U_{234}U_{345}U_{456}\ket{+}^{\otimes 6}$.
  The local measurements in the $Z$ basis are pictured by arrows pointing in the upper right direction and the numbers $+1,-1$ denote the outcomes.
Part $a)$ of the figure shows how one can obtain a $3-$qubit graph state $\E(\proj{\Psi})$ with $\ket{\Psi}_3=U_{13}U_{34}\ket{+}^{\otimes 3}$. Part b) indicates how a bi-partite state $\E(\proj{\Psi})$ with $\ket{\Psi}_2=U_{12}\ket{+}^{\otimes 2}$ is obtained. The coloring of the qubits refer to the three colors of the LMES.}.{\label{cut}}
\end{center}
\end{figure}

Let us now explain how the indirect purification protocols work. We consider first the scenario where noisy graph states are obtained from Eq. (\ref{eq_whitenoise}) by performing local $Z$--measurements. Those states can then be recombined via LOCC to obtain the desired LME.
We consume four copies of the state given in Eq. (\ref{eq_whitenoise}) to achieve this task. Let us denote by $m_i^j$ the measurement outcome of $Z_j$ measured on the $i$--th copy. First of all, the 2nd and 5th qubit of copy $1$ and $3$ are measured in the $Z$--basis. We keep only those instances where the outcomes $m_2^1=1, m_5^1=-1$, and $m_2^3=-1, m_5^3=1$ are obtained. In the following we do not consider the qubits which have been measured, since they factor from the resulting noisy graph (bi-partite) state. The resulting state of copy $1$ ($3$) is given by  Eq. (\ref{eq_whitenoise}) with $\ket{\Psi}=U_{13}U_{34}\ket{+}^{\otimes 3}$ ($\ket{\Psi}=U_{34}U_{46}\ket{+}^{\otimes 3}$) resp. (see Figure \ref{cut} $a)$.

\begin{figure}[h!]
\begin{center}
  \includegraphics[scale=1.5]{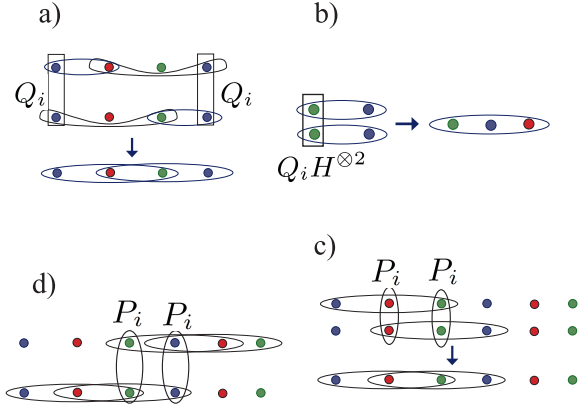}
  \caption{This figure shows how one could connect two graph states (bi-partite) states to obtain LME states $[a,(b)]$ and how one can recombine LMES $(c,d)$.
   In part a) the LMES $U_{123}U_{234} \ket{+}^{\otimes 4}$ is obtained by applying the projector $Q_{i}$ (see main text) between graph states of the form $U_{12}U_{24}\ket{+}^{\otimes 3}$ and $U_{1'3'}U_{3'4'}\ket{+}^{\otimes 3}$.
   In part b) $Q_{i}H^{\otimes 2}$ is applied to maximally entangled states $\ket{\Phi^{+}}$ resulting in $U_{123}\ket{+}^{\otimes 3}$.
   Part c) and d) show how LMES can be recombined by employing the projector $P_{i}$ (see main text).
   The coloring of the qubits refers in all parts to the colors of the quantum state.}.{\label{reconnect}}
\end{center}
\end{figure}

On the second copy qubit $3$ and $5$ are measured. The copy is kept only if $m_3^2=1$ and $m_5^1=-1$. The resulting state is then given by Eq. (\ref{eq_whitenoise}) with $\ket{\Psi}=U_{12}U_{24}\ket{+}^{\otimes 3}$. Measuring qubit $2$ and $4$ of the fourth copy and obtaining $m_2^4=-1, m_4^4=1$ leads to the state given in Eq. (\ref{eq_whitenoise}) with $\ket{\Psi}=U_{35}U_{56}\ket{+}^{\otimes 3}$.
After these measurements all graph states are purified using the purification protocol presented in \cite{DuAsBr}. Note that all states are $2$--colorable graph states. Note that the output of this process has the same fidelity with the target graph state as the original noisy LME state $\E_{wn}(\proj{\Psi})$ with the target LME state. Given the results presented in \cite{DuAsBr} it is easy to see that the protocol converges only if the original fidelity was $f\geq 0.349$. Next, we explain how these four pure graph states can be mapped to the desired LMES via LOCC. To this end let us denote by $Q_{i}$ the measurement operator $p_{i i^\prime}H\otimes H$, acting on qubit $i$ of two different states. Here, $p_{i, i^\prime}=\sqrt{2}(\ket{0}\bra{++}+\ket{1}\bra{++}U_{i, i^\prime}$, where $U_{i, i^\prime}$ denotes a $2-$qubit $\pi-$phase gate acting on the $i$--th qubit of both states and $H$ denotes the Hadamard operation. Whenever it is clear from the context we will omit the subscript. First, the states which were obtained from copy $1$ and $2$ are recombined by performing the local measurements $Q_{1}$ and $Q_{4}$ (see Figure \ref{reconnect} $a$). On copy $3$ and $4$ $Q_{i}$ for $i=3,6$ is performed. Finally, the operations $P_3$ and $P_4$ with $P_{i}=\sqrt{2}(\ket{0}_{i}\bra{00}_{ii^\prime}+\ket{1}_{i}\bra{11}_{i i^\prime})$ are performed on the two LME states which were obtained in the previous step (see Figure \ref{reconnect} $c)$). The resulting state is the desired LMES, $\ket{\Psi}$.

An other approach is to distill pure bi--partite maximally entangled states and to use them to generate the LMES via LOCC. In order to get a noisy two qubit state one could measure $4$ consecutive qubits in the $Z$--basis obtaining the outcomes $+1,-1,-1,-1$ as illustrated in Figure \ref{cut} $b)$.
In order to purify maximally entangled two--qubit states an initial fidelity of $f\geq 0.5$ is required \cite{BeBr}. Once the bipartite states are purified one can recombine them again to obtain the LMES in the following way. We use two of the maximally entangled states, say $\ket{\Phi^+}_{12}$ and $\ket{\Phi^+}_{2^\prime3}$ and perform the local projector, $Q_{2}H^{\otimes 2}$ to obtain the $3-$qubit LME state
$Q_{2} H^{\otimes 2} \ket{\Phi^{+}}_{12}\ket{\Phi^{+}}_{2^\prime 3}=U_{123}\ket{+}^{\otimes 3}$ (see Figure \ref{reconnect}).
$\ket{\Psi}_6$ is then obtained by taking four of those $3-$qubit LME states and connecting them in a similar way. For instance, to obtain $U_{i jk}U_{jkl^\prime}\ket{+}^{\otimes 6}$ for some $i,j,k,l^\prime \in V$ we perform the measurements on qubits $j,j^\prime$ and on qubits $k,k^\prime$, i.e. $P_{j} P_{k} U_{i j k}U_{j^\prime k^\prime l^\prime}\ket{+}^{\otimes 6}= U_{i jk}U_{jkl^\prime}\ket{+}^{\otimes 6}$ (see Sec. \ref{SecPPLME} and Figure \ref{reconnect}).

\section{Conclusion}
\label{seccon}

In this paper we have devised a multipartite entanglement purification protocol for LME states with $\pi$-phase interactions. These constitute a large class of states with interesting physical and mathematical properties, which includes and generalizes well-known subclasses such as stabilizer states and graph states. In particular, we have proposed a purification procedure for the subclass of regular LME states which, in turn, can be used as a building block for protocols in which the target state is not regular. The main idea behind the protocol traces back to the Bell-state recurrence protocol: two copies of the noisy state are used to obtain some information about the eigenvalues of the stabilizers which is revealed by measuring (i.\ e.\ destroying) just one copy. However, LME states have very different mathematical properties (e.\ g.\ the stabilizers are not local) and have more complex interaction patterns (which affects the way in which quantum correlations are spread through the state) and it has been necessary to develop significatively different tools and techniques to achieve this task. Another difference with previous purification protocols is that, as we have proven, depolarizing to an LMES-diagonal form is in general not possible by deterministic LOCC. However, we have shown that depolarization is not necessarily advantageous for the success of purification protocols. We have also considered different examples to illustrate the performance of our protocol, determining the maximum amount of local noise under depolarizing and dephasing channels and the minimal fidelity with the target state under global white noise that allows for the convergence of the purification protocol. Remarkably, we have additionally shown that our protocol purifying directly to the LME state has a better tolerance than indirect strategies that purify to several different graph states from which the LME state can be obtained locally (i.\ e.\ our protocol converges for states which are too noisy for the other strategies).

It is worth pointing out that by convergence of our protocol we mean that the fidelity of the noisy state with the target state can be brought arbitrarily close to 1 by repeating our scheme sufficiently often. However, the fidelity only reaches 1 in the limit of infinitely many iterations. This means that, strictly speaking, the yield $Y$ of our protocol, which is defined as the ratio $m/n$ between the number $m$ of perfect copies of the target state obtained from $n$ copies of the noisy state when $n$ goes to infinity, is zero. This is also true for previously proposed purification protocols for graph states \cite{AsDuBr, KrDu} and is not very relevant from the practical point of view. However, from the theoretical point of view it is interesting to note that these protocols can be modified using the hashing and breeding schemes \cite{BeBr, BeWo} to obtain a non-zero yield for states with very large fidelity with the target state. The idea behind these protocols is to regard the $n$ copies of the depolarized noisy state $\rho^{\otimes n}=\sum_{\k^{(1)}\cdots\k^{(n)}}\lambda_{\k^{(1)}}\cdots\lambda_{\k^{(n)}}|\Psi_{\k^{(1)}}\rangle\langle\Psi_{\k^{(1)}}|\otimes\cdots\otimes|\Psi_{\k^{(n)}}\rangle\langle\Psi_{\k^{(n)}}|$ as a statistical mixture. One then needs to identify the bit string $\vec{\k}=(\k^{(1)},\ldots,\k^{(n)})$ that was actually produced by the source taking into account that the probability of emitting each of the bits in $\vec{\k}$ is encoded in $\lambda_\k$. Once $\vec{\k}$ is determined the state is identified to be $|\Psi_{\k^{(1)}}\rangle\otimes\cdots\otimes|\Psi_{\k^{(n)}}\rangle$, which can be transformed by local unitaries to $|\Psi_\textbf{0}\rangle^{\otimes n}$. Let us consider, for instance, the first bit of $\k$, $k_{a_1}$, which is 0 with probability $p(k_{a_1}=0)=\sum_{\k\setminus k_{a_1}}\lambda_{0k_{a_2}\cdots k_{a_{I_A}},\k_{\bar{A}}}$ and 1 with probability $p(k_{a_1}=1)=\sum_{\k\setminus k_{a_1}}\lambda_{1k_{a_2}\cdots k_{a_{I_A}},\k_{\bar{A}}}$. Since the protocols allow to read the parity of the bits, in the asymptotic limit of many copies one just needs to measure (i.\ e.\ waste) $nS(p_{a_1})$ copies of the state to determine the first substring of $\vec{\k}$, $\vec{\k}_{a_1}=(k_{a_1}^{(1)},\ldots,k_{a_1}^{(n)})$. Here, $S(p_{a_1})$ is the Shannon entropy of the probability distribution $\{p(k_{a_1}=0),p(k_{a_1}=1)\}$. Since the parity checks of bits in the same color can be done jointly this protocol yields $Y=1-\max_iS(p_{a_i})-\max_iS(p_{b_i})-\cdots$, which can be strictly larger than zero. Notice that for the purification of LME states one can also read jointly the parity of bits in the same color. However, in our protocol this is achieved probabilistically by doing several GHZ-projections. In general, the probability of implementing each of these projections successfully is strictly less than one. Therefore, a correct reading of the parity of some substring of bits requires exponentially many copies, i.\ e.\ scales exponentially with $n$, thus rendering impossible a modification as above of our protocol to obtain a non-zero yield.

\section{Acknowledgment}
The research was funded by the Austrian Science Fund (FWF): Y535-N16, P20748-N16, P24273-N16 and
F40-FoQus F4011/12-N16.

\end{document}